\documentclass[9pt,twocolumn,twoside]{opticajnl}
\journal{opticajournal} 

\setboolean{shortarticle}{true}

\usepackage{amsmath,amssymb}
\usepackage[justification=justified,format=plain]{caption}

\title{Experimental photon addition and subtraction in multi-mode and entangled optical fields}

\author[1,*]{Kishore Thapliyal}
\author[1]{Jan Pe\v{r}ina Jr.}
\author[1]{Ond\v{r}ej Haderka}
\author[2]{ V\' aclav Mich\' alek}
\author[2]{Radek Machulka}
\affil[1]{Joint Laboratory of Optics, Faculty of Science, Palack\'{y} University,
Czech Republic, 17. listopadu 12, 779~00 Olomouc, Czech Republic}
\affil[2]{Joint Laboratory of Optics of Palack\'{y} University and
Institute of Physics of the Czech Academy of Sciences, Institute
of Physics of the Czech Academy of Sciences, 17. listopadu
1154/50a, 779 00 Olomouc, Czech Republic}
\affil[*]{kishore.thapliyal@upol.cz}

\begin{abstract}
Multiple photon addition and subtraction applied to multi-mode
thermal and sub-Poissonian fields as well as twin beams is
mutually compared using one experimental setup. Twin beams with
tight spatial correlations detected by an intensified CCD camera
with high spatial resolution are used to prepare the initial
fields. Up to three photons are added or subtracted to arrive at
the nonclassical and non-Gaussian states. Only the
photon-subtracted thermal states remain classical. In general, the
experimental photon-added states exhibit greater nonclassicality
and non-Gaussianity than the comparable photon-subtracted states.
Once photons are added or subtracted in twin beams, both processes
result in comparable properties of the obtained states owing to
twin-beam photon pairing.
\end{abstract}

\setboolean{displaycopyright}{false} 

\begin{document}

\maketitle

Thermal states (TSs) of light were the pertinent quantum states in
the pioneering theoretical and experimental quantum optics.
Hanbury-Brown and Twiss, using the thermal light, revealed the
photon-number correlations characterizing bunching of photons
\cite{Brown1956}. They have been widely used in ghost imaging
\cite{Gatti2004},   
Hong-Ou-Mandel-like
interference \cite{Liu2013}, and super-resolution imaging with
 multi-photon interference \cite{Classen2016}. 

Unlike the super-Poissonian TSs, with their photon-number
fluctuations exceeding the classical limit of the Poissonian
states, sub-Poissonian states (SPSs) are endowed with
photon-number fluctuations suppressed below the Poissonian
classical limit. For this reason they have gathered significant
interest due to their nonclassical properties
\cite{Davidovich1996,Lamperti2014}. These states can emerge in numerous
processes including resonance fluorescence, squeezed light   
generation, 
light-emission from diodes,  
or Franck-Hertz experiment (for details, see \cite{PerinaJr2013b}).              

Twin beams (TWBs) composed of photon pairs with thermal
photon-number statistics 
are
the most common representatives of the large group of fields
endowed with quantum correlations called entanglement. These
correlations originate in photon pairing that is the natural
property of the nonlinear process of spontaneous parametric
down-conversion \cite{Mandel1995}. Their quantum photon-number
correlations allowed to rule out neoclassical physical theories
\cite{Giustina2015}, demonstrate quantum nonlocality
\cite{Franson1999}, or perform teleportation
\cite{Bouwmeester1997} of the state of a photon. They have been
found useful in metrology \cite{Klyshko1980}, correlated imaging
\cite{Genovese2016} as well as quantum communications
\cite{Xu2020} and computation                           
\cite{Braunstein2005b}.                         

The marginal beams of a TWB are in TSs. Similarly, using
post-selection \cite{Laurat2003,PerinaJr2013b} based on the
measurement of a given number of photons in one beam of the TWB,
we arrive at a SPS in the remaining beam. This then allows to
generate all three kinds of states, i.e. TSs, SPSs and TWBs, with
their specific properties together in one experimental setup.
Similarly, also their description can be unified (see Supplemental
document (SD) \cite{Thapliyal2024SD}). Multi-mode fields in
general do not require description in the phase space. Their
quasi-distributions of intensities provide their complete
characterization. Assigning to $j $th spatio-spectral mode of a
TWB quasi-distribution $ P_j(W_{{\rm s},j},W_{{\rm i},j}) $ of its
signal- and idler-beam intensities $ W_{{\rm s},j} $ and $ W_{{\rm
i},j} $, respectively, the quasi-distribution $ P(W_{\rm s},W_{\rm
i}) $ of the overall TWB with $ M $ modes is determined along the
formula \cite{Perina1991}:
\begin{eqnarray}   
  P(W_{\rm s},W_{\rm i}) &=& \prod_{j=1}^{M} \int_{0}^{\infty} dW_{{\rm s},j}
   \int_{0}^{\infty} dW_{{\rm i},j} \delta \left( W_{\rm s} - \sum_{k=1}^{M} W_{{\rm s},k}\right)
   \nonumber \\
  & & \times \delta \left( W_{\rm i} - \sum_{k=1}^{M} W_{{\rm i},k}\right)
  P_j(W_{{\rm s},j},W_{{\rm i},j});
\label{1}
\end{eqnarray}
$ \delta $ denotes the Dirac function. Using the Mandel detection
formula (see SD), similar relation is established among the TWB
photon-number distribution $ p^{\rm TWB}(n_{\rm s},n_{\rm i}) $
and the distributions $ p^{\rm TWB}_j(n_{{\rm s},j},n_{{\rm i},j})
$, $ j=1,\ldots,M $, of the constituting modes:
\begin{equation}   
  p^{\rm TWB}(n_{\rm s},n_{\rm i}) = \prod_{j=1}^{M} \sum_{n_{{\rm s},j},n_{{\rm i},j}=0}^{\infty}
  \delta_{n_{\rm s},\sum_{k=1}^{M} n_{{\rm s},k}}
  \delta_{n_{\rm i},\sum_{k=1}^{M} n_{{\rm i},k}}
  p^{\rm TWB}_j(n_{{\rm s},j},n_{{\rm i},j})
\label{2}
\end{equation}
and $\delta $ stands for the Kronecker symbol. Provided that {an
ideal, i.e. noiseless, TWB is composed of $ M $ equally populated
modes} with $ B $ mean photon-pair numbers, its photon-number
distribution $ p^{\rm TWB}_{\rm id}(n_{\rm s},n_{\rm i}) $ is
obtained in the form of the Mandel-Rice formula \cite{Mandel1959}:
\begin{equation}   
  p^{\rm TWB}_{\rm id}(n_{\rm s},n_{\rm i}) = \delta_{n_{\rm s},n_{\rm i}}
  \frac{\Gamma(n+M)}{n! \Gamma(M)} \frac{B^n}{(1+B)^{n+M}}
\label{3}
\end{equation}
using the $ \Gamma $ function.

Properties of these states can significantly (qualitatively) be
modified using the processes of photon subtraction (PS)
\cite{Agarwal1992b} 
and photon
addition (PA) \cite{Agarwal1991,Agarwal1992}. These processes
allow to generate nonclassicality in optical fields. For example,
addition of one photon into an optical field leads to the zero
probability of detecting the vacuum state and this results in
negative values of the Glauber-Sudarshan quasi-distribution
\cite{Escher2004}, which certifies the state nonclassicality. PA
increases the mean photon number while PS gradually increases (decreases)
the mean photon number in case of the original super-Poissonian
(sub-Poissonian) field \cite{Barnett2018}. 
Advantages of PA over PS have been discussed in~\cite{Zhang2013}.
Experimental PS from a single-mode TWB (two-mode squeezed vacuum
state) \cite{MaganaLoaiza2019}, multi-mode TWB
\cite{Thapliyal2024,PerinaJr2024}, and multi-mode graph states
\cite{Ra2020,Walschaers2020} have been performed.
{The PA and PS TWBs are endowed with better properties compared to TWBs which is appealing for numerous applications including quantum metrology. They can find their applications in entangled-photon virtual-state spectroscopy \cite{Thapliyal2024,PerinaJr2024a} and quantum imaging \cite{Genovese2016}.}

The above mentioned applications of TSs, SPSs, and TWBs pose the
question about efficiency in modifying their properties, namely
their amount of nonclassicality, applying the processes of in
general multiple PS and PA. The answer can be addressed from two
different points of view: the theoretical one and the experimental
one. Whereas the former point of view can be addressed with the
help of existing literature and SD, the latter one is elaborated
in this letter. To reach comparable experimental conditions for
this comparison, we have suggested and built an experimental setup
in which we use two corresponding portions of the emission cone of
spontaneous parametric down-conversion from a nonlinear crystal to
realize five fields with mutual correlations such that suitable
reductions or post-selections allow to built all the states
discussed above.

\begin{figure}[t]   
  \centerline{\includegraphics[width=0.99\hsize]{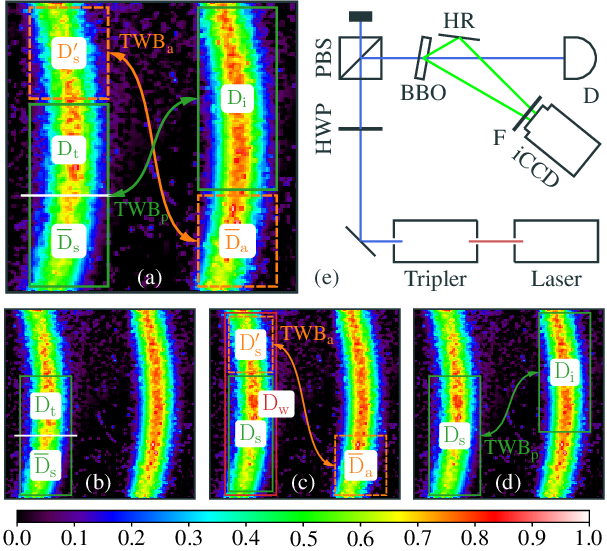} }
 \caption{(a) Five detection areas $ {\rm D}'_{\rm s} $, $ {\rm D}_{\rm t} $, $\bar{\rm
  D}_{\rm s} $, $ \bar{\rm D}_{\rm a} $, and $ {\rm D}_{\rm i} $,
  functioning as photon-number-resolving detectors, are identified on the photocathode of an iCCD camera.
  The fields detected in these areas are divided into two twin beams TWB$ {}_{\rm p,a} $.
  Photon-subtracted (added) signal beams of TWB$ {}_{\rm p} $ are
  detected in $ {\rm D}_{\rm t} $ ($ {\rm D}_{\rm w} \equiv {\rm D}_{\rm t}+\bar{\rm
  D}_{\rm s}+{\rm D}'_{\rm s} $) after post-selection in $\bar{\rm
  D}_{\rm s} $ ($ \bar{\rm D}_{\rm a} $), while the idler beam is monitored in $ {\rm D}_{\rm i} $.
  Photon pairing and the ensuing tight spatial correlations (the area of intensity cross-correlations is much smaller
  than the DAs) are indicated by arrows.
  (b) PSTSs are measured in $ {\rm D}_{\rm t} $
  after post-selecting detection in $\bar{\rm D}_{\rm s} $.
  (c) PATSs are characterized in $ {\rm D}_{\rm w}$ after post-selection on TWB$ {}_{\rm a} $
   in $\bar{\rm D}_{\rm a} $.
  (d) SPSs are observed in $ {\rm D}_{\rm s} \equiv {\rm D}_{\rm t}+\bar{\rm
   D}_{\rm s}$ after post-selection on TWB$ {}_{\rm p} $ in
   ${\rm D}_{\rm i} $. (e) Experimental setup: TWBs are generated by actively stabilized
   (using half-wave plate HWP, polarizing beam splitter PBS, and active feedback detector D)
   ultra-short-third-harmonic pulses in nonlinear crystal BBO. Both beams are spectrally filtered
   by bandpass interference filter F before being detected by an iCCD
   camera. The idler beam is reflected by highly
   reflecting mirror HR. See the text {and SD} for more details. }
\label{fig1}
\end{figure}

The suggested experimental setup is summarized in Fig.~\ref{fig1}.
The fields were generated in a 5-mm-long type-I beta-barium-borate
(BaB$ {}_2 $O$ {}_4 $, BBO) crystal cut for a slightly
non-collinear geometry by the third-harmonic pulses (280~nm) from
a femtosecond cavity-dumped Ti:sapphire laser (840~nm, 150~fs). A
14-nm-wide bandpass interference filter was used to filter the
nearly-frequency-degenerate (at $ \approx $560~nm) down-converted
signal and idler beams. Five fields were detected in five
detection areas (DAs) on a multichannel iCCD camera Andor
DH334-18U-63 [as shown in Fig.~\ref{fig1}(a)] that are denoted as
${\rm D}'_{\rm s} $, $ {\rm D}_{\rm t} $, $ \bar{\rm D}_{\rm s} $,
$ {\rm D}_{\rm i}$, and $ \bar{\rm D}_{\rm a} $. The five fields
compose two TWBs. The DAs ${\rm D}'_{\rm s}$ and $ \bar{\rm
D}_{\rm a} $ monitor the signal and idler beams of TWB$ {}_{\rm a}
$. Similarly, the signal and idler beams of TWB$ {}_{\rm p} $ are
collected in DAs ${\rm D}_{\rm t}+\bar{\rm D}_{\rm s}\equiv {\rm
D}_{\rm s}$ and ${\rm D}_{\rm i}$, respectively.

\begin{table}  
 {\setlength{\tabcolsep}{3pt}
   \small
 \begin{tabular}{p{1.5cm}  p{1.5cm}  p{1.5cm}  p{1.7cm}  p{1.5cm}}
 
  \hline
  State    & Detection area (DA) & Frequency histogram & Photon number distribution & Remarks  \\
 \hline
  $\bar{c}_{\rm s}$-PSTS & $ {\rm D}_{\rm t};\, \bar{\rm D}_{\rm s} $ & $f(c_{\rm t};\bar{c}_{\rm s}) $ & $p^{\rm s}_{\rm th}(n) $ & Fig.~\ref{fig1}(b)  \\
  $\bar{c}_{\rm a}$-PATS/ & $ {\rm D}_{\rm w};\, \bar{\rm D}_{\rm a}$ & $f(c_{\rm w};\bar{c}_{\rm a}) $ & $p^{\rm a}_{\rm th}(n) $ & Fig.~\ref{fig1}(c)  \\
  $\bar{c}_{\rm a}$-PATSsc & & & & \\
  $\bar{c}_{\rm s}$-PSSPS & $ {\rm D}_{\rm t};\, \bar{\rm D}_{\rm s}, \, {\rm D}_{\rm i}$ & $f(c_{\rm t};\bar{c}_{\rm s},c_{\rm i}) $ & $p_{\rm sP}^{\rm s}(n)$ & - \\
  {\small $\bar{c}_{\rm a}$-PASPS/} & $ {\rm D}_{\rm w};\, \bar{\rm D}_{\rm a}, \, {\rm D}_{\rm i}$ & $f(c_{\rm w};\bar{c}_{\rm a},c_{\rm i}) $ & $p^{\rm a}_{\rm sP}(n) $ & - \\
  {\small $\bar{c}_{\rm a}$-PASPSsc} & & & & \\
  $\bar{c}_{\rm s}$-PSTWB & $ {\rm D}_{\rm t}, \, {\rm D}_{\rm i};\, \bar{\rm D}_{\rm s}$ & $f(c_{\rm t},c_{\rm i};\bar{c}_{\rm s}) $ & $p^{\rm s}_{\rm TWB}(n_{\rm s},n_{\rm i})$ & Fig.~\ref{fig1}(a)  \\
  $\bar{c}_{\rm a}$-PATWB & $ {\rm D}_{\rm w}, \, {\rm D}_{\rm i};\, \bar{\rm D}_{\rm a}$ & $f(c_{\rm w},c_{\rm i};\bar{c}_{\rm a}) $ & $p^{\rm a}_{\rm TWB}(n_{\rm s},n_{\rm i})$ & Fig.~\ref{fig1}(a)  \\
  \hline
 \end{tabular}}
 \centering \caption{{\bf Characteristics of {PSTS (PATS): {\rm Photon Subtracted (Added) Thermal State,} PSSPS (PASPS): {\rm Photon Subtracted (Added) Sub-Poissonian State,} PSTWB (PATWB): {\rm Photon Subtracted (Added) TWin Beam} and PATSsc (PASPSsc) {\rm using spatial correlations}}}; photocount numbers in
  DAs ${\rm D}_{\rm s}\equiv {\rm
  D}_{\rm t}+\bar{\rm D}_{\rm s} $ and ${\rm D}_{\rm w}\equiv {\rm
  D}_{\rm s}+{\rm D}'_{\rm s} $ are denoted as $c_{\rm s}$ and $c_{\rm w}$, respectively.}
 \label{tab1}
\end{table}

Using the above DAs, we built the photocount histograms of the
following states obtained by subtracting (adding) $\bar{c}_{\rm
s}$ ($\bar{c}_{\rm a}$) photocounts: $\bar{c}_{\rm s}$
Photon-Subtracted Thermal States ($\bar{c}_{\rm s}$-PSTSs),
$\bar{c}_{\rm a}$ Photon-Added Thermal States ($\bar{c}_{\rm
a}$-PATSs), $\bar{c}_{\rm s}$ Photon-Subtracted Sub-Poissonian
States ($\bar{c}_{\rm s}$-PSSPSs), $\bar{c}_{\rm a}$ Photon-Added
Sub-Poissonian States ($\bar{c}_{\rm a}$-PASPSs), $\bar{c}_{\rm
s}$ signal Photon-Subtracted TWin Beams ($\bar{c}_{\rm
s}$-PSTWBs), and $\bar{c}_{\rm a}$ signal Photon-Added TWin Beams
($\bar{c}_{\rm a}$-PSTWBs). Also, considering the tight spatial
photon-pair correlations in TWB$ _{\rm a} $
\cite{Hamar2010,PerinaJr2017}, we generated the $\bar{c}_{\rm
a}$-PATSs and $\bar{c}_{\rm a}$-PASPSs with improved properties, {and we labelled them by
sc:} $\bar{c}_{\rm a}$-PATSscs and $\bar{c}_{\rm a}$-PASPSscs.
Characteristics of all these states are summarized in
Tab.~\ref{tab1}.

The frequency histograms of the original fields are obtained as
follows. The frequency histogram $f(c_{\rm s}) $ of a multi-mode
TS arises from $ {\rm D}_{\rm s} $, while those of TWB$_{\rm
p}$ [$f(c_{\rm s},c_{\rm i}) $] and SPS [$f(c_{\rm s};c_{\rm i})
$] are reached using $ {\rm D}_{\rm s}$ and ${\rm D}_{\rm i}$.
A symmetric beam splitter in the signal beam of TWB${}_{\rm p} $,
needed for PS, is realized by splitting the DA $ {\rm D}_{\rm s} $
into two DAs $ {\rm D}_{\rm t} $ and $ \bar{\rm D}_{\rm s} $,
where the latter serves for counting the post-selecting
photocounts. On the other hand, the fields for PA occur in $
{\rm D}'_{\rm s} $ provided that $\bar{c}_{\rm a}$ photocounts is
registered in $ \bar{\rm D}_{\rm a} $. Exploiting the tight
spatial correlations in TWB$ {}_{\rm a} $, as observed in DAs $
{\rm D}'_{\rm s} $ and $ \bar{\rm D}_{\rm a} $, we mimic the
behavior of an ideal photon-number-resolving detector in the
post-selecting DA $ \bar{\rm D}_{\rm a} $. This results in general
in considerable improvement of nonclassical properties of PA
fields.

Using the method of detector calibration by TWBs
\cite{PerinaJr2012a} and independent measurements, parameters of
the DAs were determined (for details, see SD). They include the
detection efficiencies belonging to the introduced DAs: $
\eta_{\rm t}=\eta'_{\rm s}=\bar{\eta}_{\rm s} = 0.234\pm 0.005 $
and $\eta_{\rm i}=\bar{\eta}_{\rm a} = 0.227\pm 0.005 $. The
calibration method also provided the parameters of twin beams TWB$
_{\rm p} $ and TWB$ _{\rm a} $ in the form of multi-mode Gaussian
fields composed of photon-pair, noise-signal, and noise-idler
components (for details and parameters, see SD). They are used to
determine the theoretical values in the graphs. On the other hand,
the maximum-likelihood reconstruction
\cite{Dempster1977,PerinaJr2012} was applied to arrive at the
photon-number distributions corresponding to the experimental
photocount histograms of the analyzed fields (see the third and
the fourth columns of Tab.~\ref{tab1}).

To study PS and PA, we experimentally generated 3 fields of
comparable intensities: multi-mode TS with mean photon number $
\bar{n}_{\rm th} = 5.470\pm 0.002 $ ($M_{\rm th}= 46.8\pm 0.3$,
$B_{\rm th}=0.120\pm 0.008$), SPS with mean photon number $
\bar{n}_{\rm sP} = 6.030\pm 0.005 $ (arising after detecting
$c_{\rm i}=2$ photocounts in D$ _{\rm i} $), and TWB with mean
photon-pair number $ \langle n_{\rm p}^{\rm p}\rangle = 4.85\pm
0.05 $ (for details, see SD).

First, we pay attention to PS and PA in one-beam fields TS and
SPS. They qualitatively differ: Whereas the TS is classical with
the Fano factor $ F^{\rm th} = 1.12\pm 0.01 $, the SPS is
nonclassical owing to its suppressed photon-number fluctuations
characterized by $ F^{\rm sP} = 0.84\pm 0.02 $. We note that the
Fano factor $ F $ \cite{Mandel1959},
\begin{equation}  
 F_{n} =  \langle (\Delta n)^2\rangle / \langle n\rangle,
\label{4}
\end{equation}
quantifies the magnitude of classical ($ F\ge 1 $) and quantum ($
F < 1 $, sub-Poissonian) photon-number fluctuations $ \Delta n = n
- \langle n \rangle $. Both PS and PA are probabilistic processes
and so we characterize them by the probabilities $ p^{\rm p} $ of
field generation (see SD for details). Considering the states
obtained by detecting up to two photocounts, these probabilities
are greater than 10\%. Only when the spatial photon-pair
correlations are taken into account in preparing the field to be
added, we may use the fields arising after detecting up to one
photocount to keep reasonable values of the probabilities $ p^{\rm
p} $. Applying the spatial photon-pair correlations, we generated
the states PATSscs and PASPSscs.

\begin{figure}[t]   
 \centerline{\includegraphics[width=0.32\hsize]{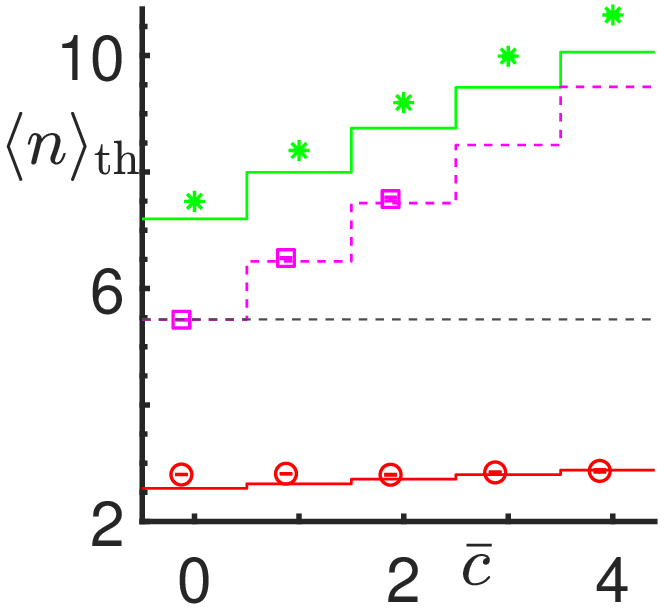}
 \includegraphics[width=0.32\hsize]{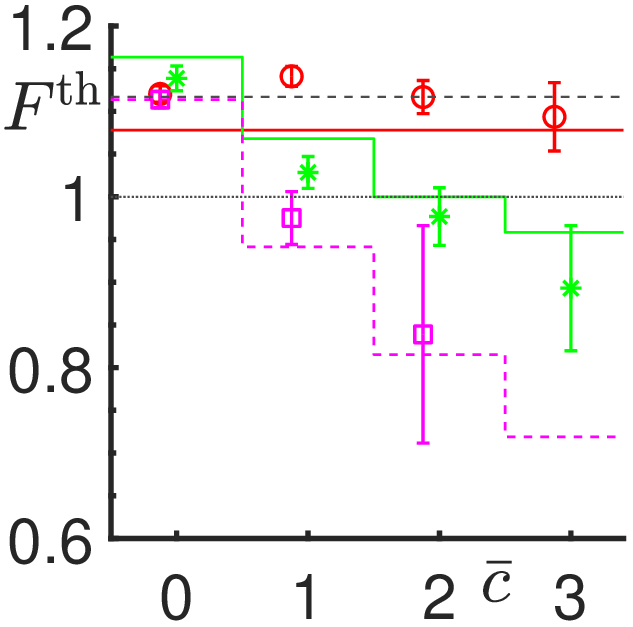}
 \includegraphics[width=0.32\hsize]{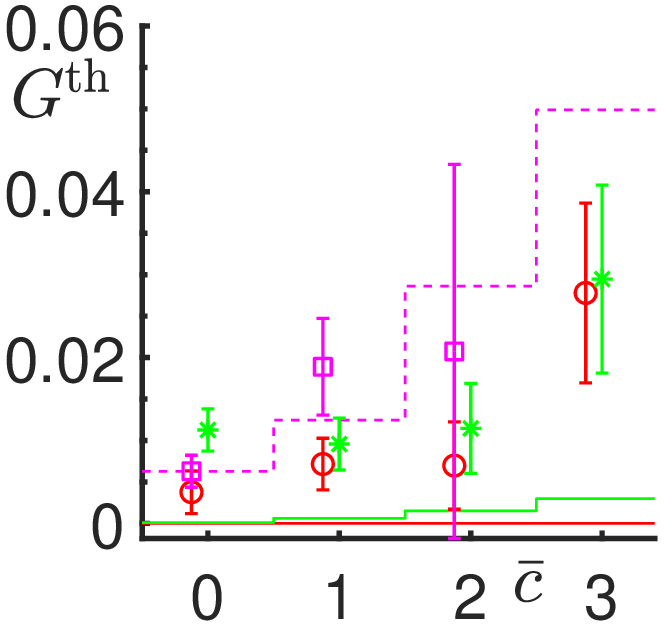}
 }
 \centerline{ \small Original thermal states \hspace{2mm} (a) \hspace{.2\hsize} (b) \hspace{.2\hsize} (c)}
  \centerline{ \includegraphics[width=0.32\hsize]{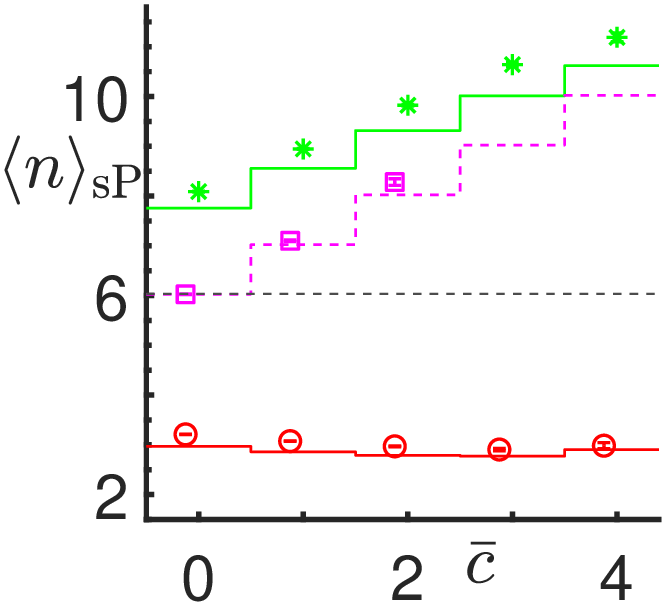}
    \includegraphics[width=0.32\hsize]{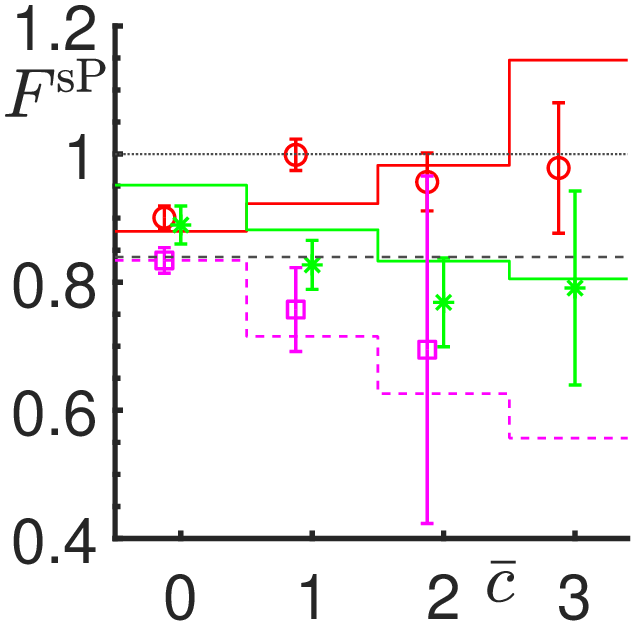}
    \includegraphics[width=0.32\hsize]{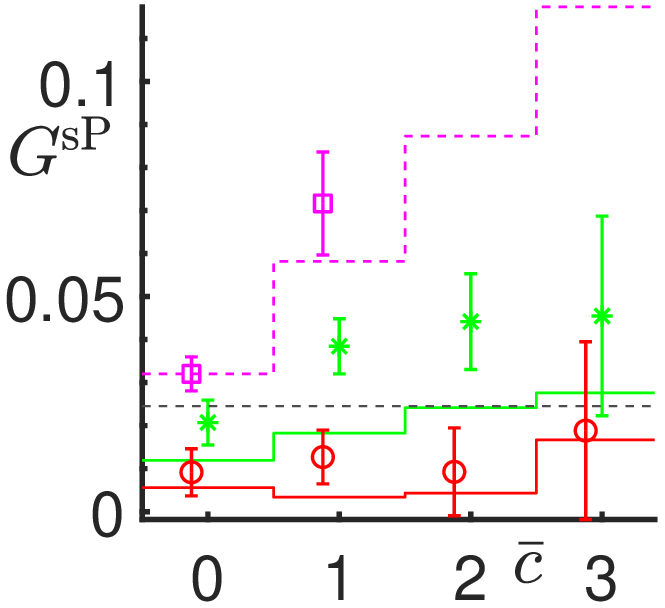}}
 \centerline{ \small Original sub-Poissonian states \hspace{2mm}  (d) \hspace{.15\hsize} (e) \hspace{.15\hsize} (f)}

 \caption{(a) [(d)] Mean photon number $ \langle n\rangle_{\rm th} $ [$ \langle n\rangle_{\rm sP} $],
  (b) [(e)] Fano factor $ F^{\rm th}$ [$ F^{\rm sP}$], and
  (c) [(f)] mutual entropy $ G^{\rm th} $ [$ G^{\rm sP} $] in PSTS and PATS [PSSPS and PASPS] as they depend
  on number $ \bar{c} $ of subtracted or added photocounts. Data for states with PS, PA, and PA with considered photon-pair spatial correlations
  are plotted in turn by red $ \circ $, green $ \ast $, and magenta $\scriptscriptstyle{\square}$.
  Experimental data are plotted as isolated symbols with error bars, solid (dashed) curves originate in the Gaussian model (and take into
  account spatial photon-pair correlations) - see SD. Statistical errors are derived from $ 1.2 \times 10^6 $ measurement repetitions.
  In (b,e), the quantum-classical border $ F = 1 $ is  plotted as a dotted black horizontal line. The dashed lines correspond to
  the values characterizing the original TS and SPS.}
\label{fig2}
\end{figure}

Interestingly, PS increases the average photon number $ \langle
n\rangle_{\rm th} $ in case of TSs [see Fig.~\ref{fig2}(a)] with the
increasing number $ \bar{c} $ of subtracted photocounts while it
ideally keeps the Fano factor $ F^{\rm th}$ unchanged [see
Fig.~\ref{fig2}(b)]. This behavior originates in natural photon
bunching of TSs \cite{Mandel1995}. Contrary to this, PS reduces
the average photon number $ \langle n\rangle_{\rm sP} $ in SPS 
with the increasing $ \bar{c} $ [see Fig.~\ref{fig2}(d)],
{e.g. we have $ \langle
n\rangle_{\rm sP}=  3.07\pm 0.01 \left(2.97\pm 0.01 \right)$ for 1- (2-)PSSPS,}
and it simultaneously increases the Fano factor $ F^{\rm sP}$ [see
Fig.~\ref{fig2}(e)].

Compared to this, PA acts qualitatively the same on both TSs and
SPSs. Their mean photon numbers $ \langle n\rangle_{\rm th} $ and
$ \langle n\rangle_{\rm sP} $ increase with the increasing number
$ \bar{c} $ of added photocounts while their Fano factors $ F^{\rm
th}$ and $ F^{\rm sP}$ gradually drop down, as documented in
Figs.~\ref{fig2}(a,d) and \ref{fig2}(b,e). The comparison of
curves in these figures reveal that the consideration of
photon-pair spatial correlations reduces mean photon numbers $
\langle n\rangle_{\rm th} $ and $ \langle n\rangle_{\rm sP} $ in
the analyzed states, but it also significantly reduces the Fano
factors $ F^{\rm th}$ and $ F^{\rm sP}$. In fact, utilization of
the spatial correlations allows to realize nearly ideal PA in TS
and SPS. Such PA allowed us to generate 1-PATSsc with $ F^{\rm th}
= 0.97\pm 0.03 $ and 2-PATSsc with $ F^{\rm th} = 0.83\pm 0.13 $.
Similarly, we observed 1-PASPSsc with $ F^{\rm sP} =0.76\pm 0.06 $
and 2-PASPSsc with $ F^{\rm sP} =0.69\pm 0.27 $.

PS and PA also modify non-Gaussianity
\cite{Allevi2010b,Ghiu2014,PerinaJr2024a} of the   
state. Non-Gaussianity is quantified by the relative entropy $ G $
\cite{Marian2013} between the field photon-number distribution $
p(n) $ and its Gaussian reference that we consider in the form of
the Mandel-Rice distribution $ p^{\rm MR}(n) $ \cite{Perina1991}
with the same mean photon number:
\begin{equation}   
 G = \sum_{n=0}^{\infty}\left[ p(n)\ln[p(n)] -
  p(n)\ln[p^{\rm MR}(n)] \right].
\label{5}
\end{equation}
According to the curves in Fig.~\ref{fig2}(c,f), only PS from SPS
results in the decrease of mutual entropy $ G^{\rm sP} $ of the
original SPS. In all other cases, PS as well as PA increase the
mutual entropies $ G^{\rm th} $ and $ G^{\rm sP} $ of the original
states: The greater the number $ \bar{c} $ of subtracted or added
photocounts is, the greater the change of the mutual entropy $ G $
is. The largest increase of mutual entropy $ G $ is reached when
the nearly ideal PA is performed using the spatial photon-pair
correlations. We note that the mutual entropy $ G^{\rm th} $ of
the original experimental TS in Fig.~\ref{fig2}(c) is larger than
0 which reflects the experimental imperfections that are not
concealed in the maximum-likelihood reconstruction. These
imperfections also lead to the systematically greater experimental
values of $ G $ compared to their theoretical predictions.

\begin{figure}[t]  
\begin{centering}
\includegraphics[width=0.32\hsize]{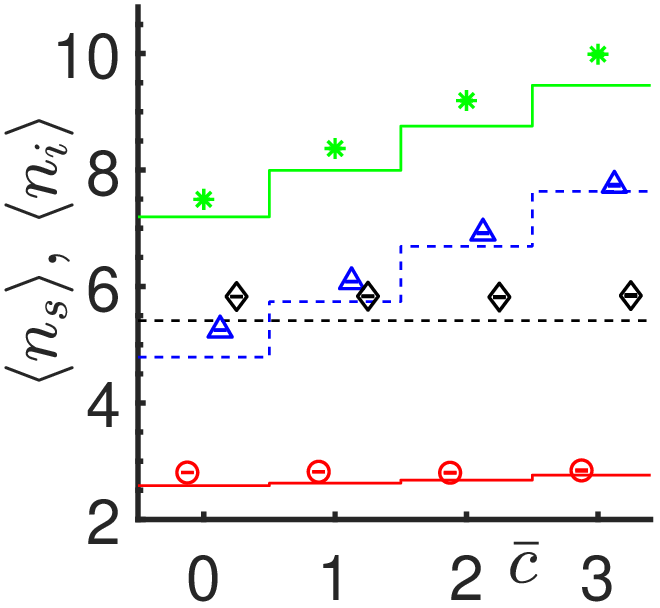}
 \includegraphics[width=0.32\hsize]{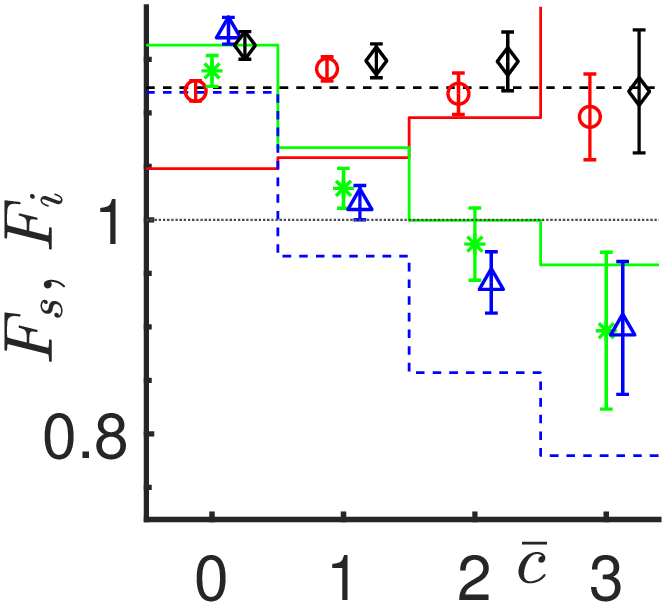}
\includegraphics[width=0.32\hsize]{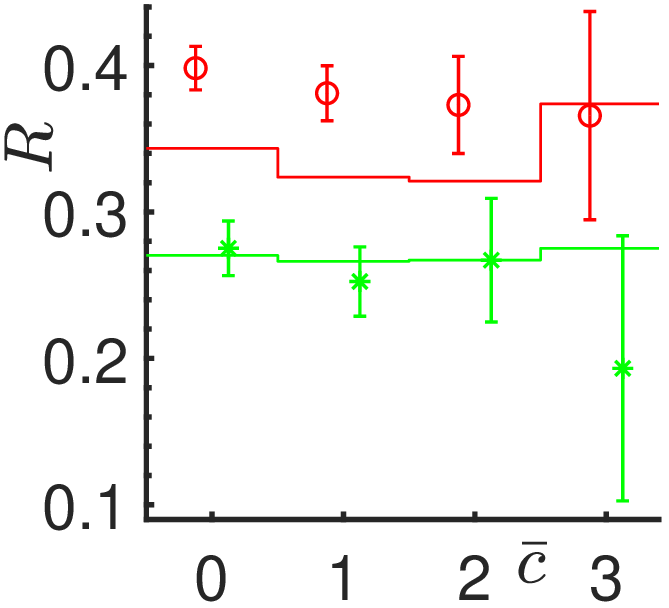}
 \centerline{ \small Original twin beam \hspace{2mm}  (a) \hspace{.2\hsize} (b) \hspace{.2\hsize} (c)}
 \par\end{centering}
\caption{(a) Mean photon numbers $ \langle n_{\rm s}\rangle $ and
  $ \langle n_{\rm i}\rangle $ and (b) Fano factors $F_{s}$
  and $F_{i}$ of the signal and idler beams, respectively, and
  (c) noise-reduction parameter $R$ as they depend on the number $ \bar{c} $
  of signal photocounts subtracted or added in TWB.
  In (a,b), for PS [PA], $ \langle n_{\rm s}\rangle $ and  $F_{s}$
  is drawn by red $ \circ $ [green $ \ast $], and $ \langle n_{\rm i}\rangle $
  and $F_{i}$ is plotted by blue $ \triangle $ [black $
  \diamond$]. In (c), $ R $ is drawn by red $ \circ $ [green $ \ast $]
  for PS [PA].
  Experimental data are plotted as isolated symbols with error bars,
  solid (signal beam) and dashed (idler) curves originate in the Gaussian model. In (b), the quantum-classical
  border $ F = 1 $ is plotted as a dotted black
  horizontal line.} \label{fig3}
\end{figure}

The differences between PS and PA, as discussed above, are
considerably reduced once we subtract or add photons into TWBs and
monitor the properties of both beams. The reason is that, due to
the photon-number correlations in a TWB, PS in the signal beam
acts as PA in the idler beam \cite{PerinaJr2024,Thapliyal2024},
though with lower intensity. This means that increasing the number
$ \bar{c} $ of photocounts subtracted from the signal beam we
increase the mean photon number $ \langle n_{\rm i}\rangle $ in
the idler beam. The mean signal photon number $ \langle n_{\rm
s}\rangle $ also increases (see the above discussion about photon
bunching in TS), but incomparably smaller [compare the curves with
red $ \circ $ and blue $ \triangle $ in Fig.~\ref{fig3}(a)]. On
the other hand, PA into the signal beam only gradually increases
its mean photon number $ \langle n_{\rm s}\rangle $ as the number
$ \bar{c} $ of added photocounts increases. It also keeps the
idler beam intact [see the curves with green $ \ast $ and black $
\diamond$ in Fig.~\ref{fig3}(a)]. Close similarity of PS and PA in
the TWB is manifested by nearly parallel curves giving $ \langle
n_{\rm s}\rangle $ by red $ \circ $ (green $ \ast $) and $ \langle
n_{\rm i}\rangle $ by black $ \diamond $ (blue $ \triangle $) and
the corresponding Fano factors $ F_{\rm s} $ and $ F_{\rm i} $
drawn in Fig.~\ref{fig3}(b).

Nevertheless, the $ \bar{c} $-PATWBs are endowed with greater mean
photon numbers in their beams compared to $ \bar{c} $-PSTWBs. This
results in greater quantum photon-number correlations in PATWBs,
as evidenced by the curves in Fig.~\ref{fig3}(c) showing the
noise-reduction parameter R,
\begin{eqnarray}  
R = \langle [\Delta (n_{\rm s} - n_{\rm i}) ]^2\rangle
    / (\langle n_{\rm s}\rangle + \langle n_{\rm i} \rangle ).
\label{6}
\end{eqnarray}
The smaller the parameter $ R $ is, the stronger the quantum
correlations are. Ideal photon pairing gives $ R=0 $.

Similarities of PS and PA states, discussed above for TSs, SPSs, and TWBs from the
point of view of their photon-number distributions, can also be 
addressed using the corresponding quasi-distributions of intensities \cite{Perina1991}. This provides an additional insight into this comparison, especially considering the nonclassicality \cite{Glauber1963} which is an important attribute of these states (for details, see SD).

In {\emph{conclusion}}, we have experimentally compared photon
subtraction and photon addition in a multi-mode thermal state,
sub-Poissonian state, and twin beam of comparable intensities.
Photon addition has been found advantageous over photon
subtraction as it allows to generate nonclassicality in
super-Poissonian thermal states and enhance nonclassicality in
sub-Poissonian states. Moreover, exploiting the spatial
photon-pair correlations, nearly ideal photon addition can be
performed. However, in twin beams, photon subtraction and addition
provide the states with comparable properties, even when analyzed
at the level of intensity quasi-distributions.

\begin{backmatter}

\bmsection{Funding} The authors acknowledge support by the project OP JAC CZ.02.01.01/00/22\_008/0004596 of the Ministry of Education, Youth, and Sports of the Czech Republic.


\bmsection{Disclosures}
The authors declare no conflicts of interest.

\bmsection{Data availability} Data underlying the results
presented in this paper are publicly available at https://doi.org/10.57680/asep.0587552.

\bmsection{Supplemental document} See Supplement 1 for supporting
content.

\end{backmatter}

\bibliography{thapliyal}


\end{document}


\maketitle

\section{Photon-addition and photon-subtraction in multi-mode thermal \\ and sub-Poissonian states}

Description of multi-mode thermal states (TSs) and sub-Poissonian
states (SPSs) can be unified when we consider the multi-mode TSs
as the marginal states of multi-mode twin beams (TWBs) and SPSs as
the states arising from TWBs in post-selection following
photon-number-resolved detection in one of the beams.

Considering an ideal TWB composed of only photon pairs and having
$ M_{\rm p} $ identical independent modes each populated by $
B_{\rm p} $ mean photon pairs, its {ideal (i.e. noiseless)} photon-number distribution $
p^{\rm TWB}_{\rm id}(n_{\rm s},n_{\rm i}) $ is expressed using the
Mandel-Rice distribution $ p^{\rm MR}(n;M_{\rm p},B_{\rm p}) $ as
follows:
\begin{equation} 
 p^{\rm TWB}_{\rm id}(n_{\rm s},n_{\rm i}) = \delta_{n_{\rm s},n_{\rm
 i}} p^{\rm MR}(n_{\rm s};M_{\rm p},B_{\rm p}) ,
\label{S1}
\end{equation}
where $ \delta $ stands for the Kronecker symbol and $ p^{\rm
MR}(n;M,B)= \Gamma(n+M)/[n! \Gamma(M)] B^n/(1+B)^{n+M} $ using the
$ \Gamma $ function.

Tracing over the idler beam, we obtain a multi-mode TS with the
mean photon number $\bar{n}_{\rm th} = M_{\rm th}B_{\rm th} $ and
photon-number distribution $ p_{\rm th} $:
\begin{equation}  
 p_{\rm th}(n) = \sum_{n_{\rm i}=0}^{\infty} p^{\rm TWB}_{\rm id}(n,n_{\rm i}) = p^{\rm
 MR}(n;M,B).
\label{S2}
\end{equation}

Detecting $ \bar{c}_{\rm sP} $ photocounts in the idler beam, we
obtain an SPS in the signal beam characterized by its
photon-number distribution $ p_{\rm sP} $:
\begin{eqnarray}  
 p_{\rm sP}(n;\bar{c}_{\rm sP}) &=& \sum_{n_{\rm i}=0}^{\infty}
   K(\bar{c}_{\rm sP},n_{\rm i};\eta) p^{\rm TWB}_{\rm id}(n,n_{\rm i}).
\label{S3}
\end{eqnarray}
In Eq.~(\ref{S3}), the detection matrix $ K $ gives the
probability of detecting $ c $ photocounts assuming $ n $ photons
impinging on a photon-number-resolving detector with efficiency
$\eta$ and zero dark counts \cite{Haderka2005a}:
\begin{eqnarray}  
 K(c,n;\eta) & = &   \left(\begin{array}{c} n\\c\end{array}\right)    
  \eta^{c} (1-\eta)^{n-c}.
\label{S4}
\end{eqnarray}
If $\eta=1$ we have $c=n$ in Eq.~(\ref{S4}) and the SPS
conditioned by detecting $ \bar{c}_{\rm sP} $ photocounts in the
idler beam is the Fock state with $ \bar{c}_{\rm sP} $ photons. We
note that the detection matrix $K$ given in Eq.~(\ref{S4}) is the
specific form of the general detection matrix $ T $ defined below
in Eq.~(\ref{S11}) for an iCCD camera appropriate for a camera
with infinitely many pixels and zero dark counts.

The Fano factor $ F^{\rm th} $, defined in Eq.~(3) in the main
text, is obtained for a multi-mode TS in the form
\begin{equation} 
 F^{\rm th}= 1 + \frac{\bar{n}_{\rm th}}{M_{\rm th}} = 1 + B_{\rm th}
 >1.
\label{S5}
\end{equation}
According to Eq.~(\ref{S5}), if $ M_{\rm th} \rightarrow \infty $,
then $ B_{\rm th} \rightarrow 0 $ and $ F^{\rm th} \rightarrow 1
$. Similarly, we have $F^{\rm sP}<1$ for a SPS {(see \cite{PerinaJr2013b} and references therein)}. When an ideal
detector is used ($\eta=1$) we even derive $F^{\rm sP} =0 $, {i.e. we have a Fock state.}

Photons from the states are subtracted with the help of a beam
splitter that usually has a high transmissivity $ t $
\cite{Barnett2018}. The reflected photons are counted by a
photon-number-resolving detector and photon-subtracted (PS) states
then arise in the transmitted field after post-selection.
Photon-number distribution $ p^{\rm s} $ of a PS state obtained
after detecting $ \bar{c}_{\rm s} $ photocounts in the reflected
field is determined along the formula
\begin{eqnarray}   
 p^{\rm s}(n;\bar{c}_{\rm s}) &=& \sum_{\bar{n}_{\rm s}=0}^{\infty}
  K(\bar{c}_{\rm s},\bar{n}_{\rm s};\eta)
  {\rm Bi}(n,n+ \bar{n}_{\rm s};t)
  p(n+\bar{n}_{\rm s}) ,
\label{S6}
\end{eqnarray}
in which we use the binomial distribution $ {\rm Bi}(n,m;t) = m! /
[n! (m-n)!] \; t^{n}(1-t)^{m-n} $.

Similarly, photon addition (PA) in a state can be performed using
a beam splitter \cite{Dakna1998}. Alternatively, optical
parametric amplification can be applied
\cite{Zavatta2004,Zavatta2005}. In the simplest scenario, we just
convolve photon-number distribution $ p $ of a state with that
corresponding to one (several) photons to be added or their
blurred approximations described by suitable SPSs. For an SPS
obtained after detecting $ \bar{c}_{\rm sP} $ idler photocounts,
i.e. with the photon-number distribution $ p_{\rm
sP}(n;\bar{c}_{\rm sP}) $ written in Eq.~(\ref{S3}), we have:
\begin{eqnarray}  
 p^{\rm a}(n;\bar{c}_{\rm sP}) &=& \sum_{n'=0}^{n}
  p(n') p_{\rm sP}(n-n';\bar{c}_{\rm sP}).
\label{S7}
\end{eqnarray}

\begin{figure}[htbp]  
\begin{centering}
 \includegraphics[width=0.48\hsize]{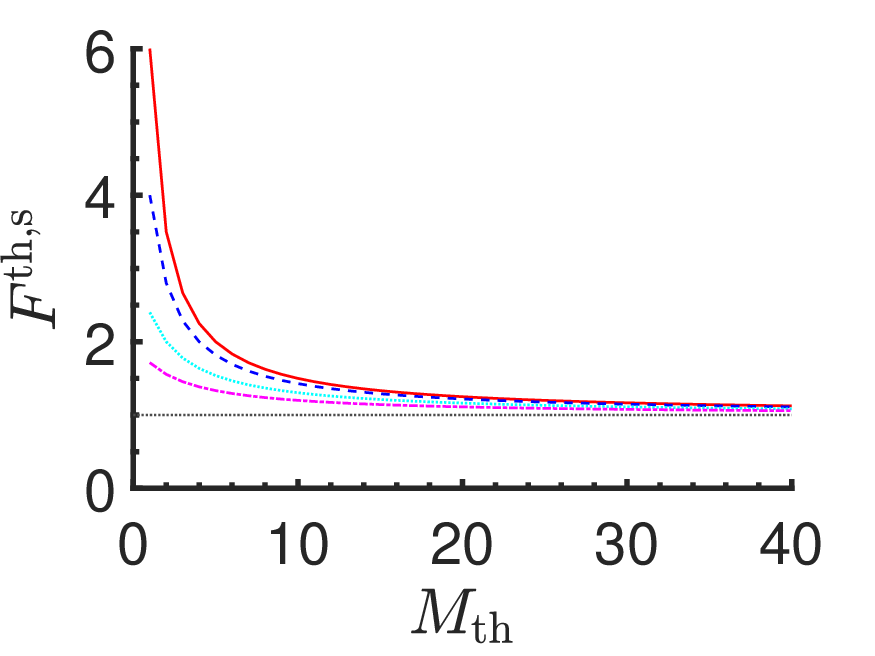}
 \includegraphics[width=0.48\hsize]{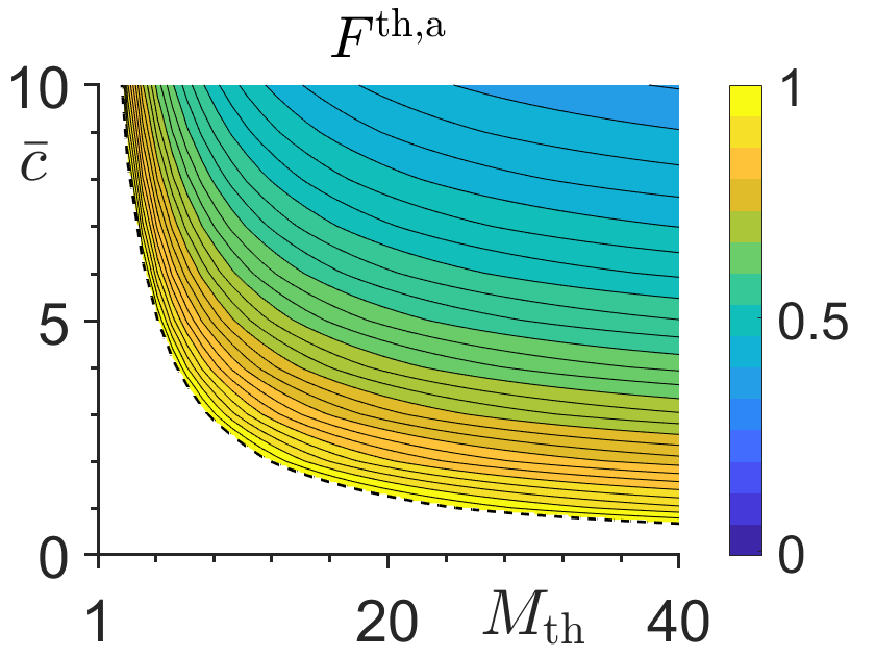}
 \centerline{ \small (a) \hspace{.45\hsize} (b)}\\
\includegraphics[width=0.48\hsize]{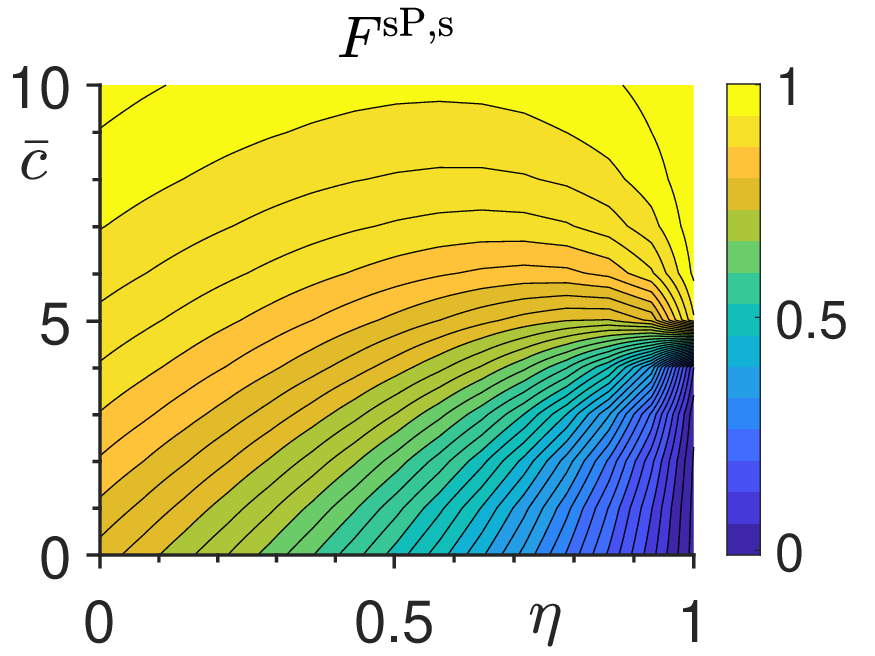}
\includegraphics[width=0.48\hsize]{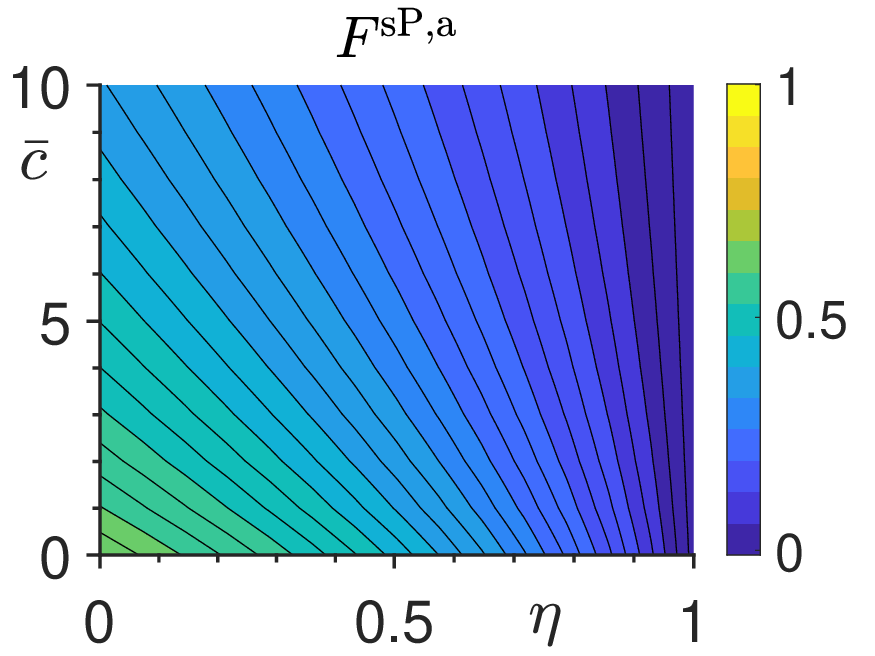}
 \centerline{ \small (c) \hspace{.45\hsize}  (d)}
 \par\end{centering}
 \caption{Fano factor (a) $ F^{\rm th,s} $ of PSTSs, (b) $ F^{\rm th,a} $ of
  PATSs, (c) $ F^{\rm sP,s} $ of PSSPSs, and (d) $ F^{\rm sP,a} $ of
  PASPSs as they depend on number $M_{\rm th}$ of
  modes/detection efficiency $ \eta $ and number $\bar{c}$ of subtracted/added photocounts
  assuming (a,b) a TS with $\bar{n}_{\rm th}= M_{\rm th} B_{\rm th} =5$ and (c,d)
  a SPS obtained by post-selecting $\bar{c}_{\rm sP}=5$
  idler photocounts in a TWB with $M_{\rm p}=50 $ and $ B_{\rm p}=0.1$.
  Beam-splitter transmissivity $t=0.5$ when subtracting photons.
  In (a), Fano factors $ F^{\rm th} $ of PSTSs do not depend on
  $\bar{c}$, and we additionally plot the curves for $t=1 $ (red
  smooth curve), 0.9 (blue dashed), 0.7 (cyan dotted), and 0.5 (magenta dot-dashed).
  The added sub-Poissonian states are prepared by post-selecting
  $\bar{c}$ photocounts in a TWB with $M_{\rm p}=50 $ and $ B_{\rm p}=0.05$.
  In (b), $ F^{\rm th} > 1 $ in the white area.
  The quantum-classical border $ F^{\rm th}= 1 $ is
  plotted in (a,b) as a dotted/dashed black curve.}
\label{figS1}
\end{figure}

Let us first consider PS from multi-mode TSs with the varying
number $M_{\rm th}$ of modes. The number $M_{\rm th}$ of modes is
an important parameter as, with the increasing $M_{\rm th}$, the
photon-number distribution of a multi-mode TS tends towards the
Poissonian one, that lies at the classical-quantum border. This
means that the greater the number $M_{\rm th}$ of modes of a TS
is, the greater the potential of this TS to generate a
nonclassical state via PS and PA is. When analyzing PS, we
consider a beam splitter with transmissivity $ t $, i.e. the
transmitted [reflected] thermal field has the mean photon number
given as $t B_{\rm th}M_{\rm th}$ [$ (1-t) B_{\rm th}M_{\rm th}$].
Using a detector with detection efficiency $ \eta $ for
post-selection, the mean photon number $ \bar{n}_{M_{\rm
th}}^{(0)} $ of the state with zero subtracted photocounts
(0-PSTS) is simply determined as a product of the mean photon
number in the transmitted field and the probability of detecting
zero photocounts in the reflected field, i.e., $\bar{n}_{M_{\rm
th}}^{(0)}= t B_{\rm th} M_{\rm th} /[ 1+ (1-t)\eta B_{\rm th} ]
$. This formula can be generalized for the case of $ \bar{c} $
subtracted photocounts ($ \bar{c} $-PSTS) (for one-mode TS, see
\cite{Barnett2018}):
\begin{equation}   
 \bar{n}_{M_{\rm th}}^{(\bar{c})}=\frac{t B_{\rm th}}{ 1 + (1-t)\eta B_{\rm th}}(M_{\rm
 th}+\bar{c}).
\label{S8}
\end{equation}
The formula in Eq.~(\ref{S8}) suggests an interesting
interpretation: Subtraction of $ \bar{c} $ photocounts formally
acts as addition of $ \bar{c} $ modes into the really existing $
M_{\rm th} $ modes with the same mean photon number
\cite{Katamadze2020,Avosopiants2021}. For $ \eta = 1 $ and in the
area around $ t \approx 1 $, we have $\bar{n}_{M_{\rm
th}}^{(\bar{c})} \approx \bar{n}_{M_{\rm th}} (1+\bar{c}/M_{\rm
th}) $. This means that the mean photon number of 1-PSTS with $
M_{\rm th} = 1 $ mode is twice of that of the initial TS ($
\bar{n}_{M_{\rm th}=1}^{(1)}\approx 2 \bar{n}_{M_{\rm th}} $).
Similarly, the Fano factor $ F_{\bar{c}}^{\rm th,s} $ of an
$\bar{c}$-PSTS is derived as
\begin{equation}  
  F_{\bar{c}}^{\rm th,s}= 1 + \frac{\bar{n}_{M_{\rm th}}^{(\bar{c})}}{M_{\rm
   th}+\bar{c} } .
\label{S9}
\end{equation}
Substituting Eq.~(\ref{S8}) into Eq.~(\ref{S9}), we conclude that
the Fano factor $F_{\bar{c}}^{\rm th,s}$ does not depend on the
number $ \bar{c} $ of subtracted photocounts. It naturally
decreases with the increasing number $ M_{\rm th} $ of modes
provided that the field mean photon number $\bar{n}_{\rm th}$ is
fixed [see Fig.~\ref{figS1}(a)]. Moreover, the Fano factor
$F_{\bar{c}}^{\rm th,s}$ also decreases with the decreasing
transmissivity $ t $, as documented in Fig.~\ref{figS1}(a). This
reflects the increased efficiency of PS with the increased mean
photon number at the post-selecting detector.

PA into multi-mode TSs is described by simple formulas provided
that we add the photons in the Fock states, which is the case of
$\eta=1$ in Eq.~(\ref{S3}). Adding $ \bar{c} $ photocounts into a
TS with mean photon number $ \bar{n}_{\rm th} $ and Fano factor $
F^{\rm th} $, we arrive at an $ \bar{c} $-PATS with mean photon
number $\bar{n}_{{\rm th},\bar{c}}^{\rm a}=\bar{n}_{\rm th}+
\bar{c} $ and Fano factor $ F_{\bar{c}}^{\rm th,a}= [ \bar{n}_{\rm
th}/(\bar{n}_{\rm th}+\bar{c})] F^{\rm th}$ \cite{Barnett2018}.
Thus, the more photocounts we add the smaller the Fano factor $
F_{\bar{c}}^{\rm th,a} $ is. For a real post-selecting detector,
according to the graph plotted in Fig.~\ref{figS1}(b), for any
number $ M_{\rm th} $ of modes of a TS, there exists the number $
\bar{c} $ of added photocounts at which sub-Poissonianity of the
field is reached.

PS from SPSs behaves qualitatively straightforwardly. The mean
photon number in an SPS decreases with the increasing number $
\bar{c} $ of subtracted photocounts \cite{Barnett2018}. Also the
Fano factor $ F_{\bar{c}}^{\rm sP,s}$ increases with the
increasing number $ \bar{c} $ of subtracted photocounts
\cite{Barnett2018}. This is documented in Fig.~\ref{figS1}(c)
where the Fano factor $ F_{\bar{c}}^{\rm sP,s}$ is plotted as a
function of detection efficiency $ \eta $ of the detector used in
preparing the original SPS. This scheme allows to tune the amount
of sub-Poissonianity in the original SPS by varying the detector
efficiency $ \eta $: The greater the efficiency $ \eta $ is, the
more sub-Poissonian the state is [see Figs.~\ref{figS1}(c,d)].

Ideal PA using the Fock states into SPSs is governed by the
formulas derived already for TSs. Thus, adding $ \bar{c} $
photocounts into an SPS with mean photon number $ \bar{n}_{\rm sP}
$ and Fano factor $ F^{\rm sP} $, we arrive at an $ \bar{c}
$-PASPS with mean photon number $\bar{n}_{{\rm sP},\bar{c}}^{\rm
a}=\bar{n}_{\rm sP}+\bar{c} $ and Fano factor $ F_{\bar{c}}^{\rm
sP,a}= [ \bar{n}_{\rm sP}/(\bar{n}_{\rm sP}+\bar{c}) ] F^{\rm sP}$
\cite{Barnett2018}. For real PA with sub-Poissonian photons added
we observe decrease of the Fano factor $ F_{\bar{c}}^{\rm sP,a} $
with each added photocount for any value of the Fano factor $
F^{\rm sP}$ of the original SPS [see Fig.~\ref{figS1}(d)].

\section{Model for photon-added and photon-subtracted states generated from twin beams}

In the experiment for generating multi-mode multiple PA and PS
states we use two independent experimental twin beams denoted as
TWB$_{\rm p} $ and TWB$_{\rm a} $. They are described by their
photon-number distributions $p^{\rm TWB}_q (n_{\rm s},n_{\rm i})$,
$ q = {\rm p,a} $. These photon-number distributions can
faithfully be described by twofold convolutions of three
Mandel-Rice distributions \cite{Perina1991} belonging to their
photon-pair, noise-signal and noise-idler components
\cite{PerinaJr2012a,PerinaJr2013a}:
\begin{equation}   
 p^{\rm TWB}_q (n_{\rm s},n_{\rm i})=  \sum_{n=0}^{\min(n_{\rm s}, n_{\rm i})}
  p^{\rm MR,s}_q(n_{\rm s}-n;M_{\rm s}^q,B_{\rm s}^q) 
 p^{\rm MR,i}_q(n_{\rm i}-n;M_{\rm i}^q,B_{\rm i}^q) p^{\rm MR,p}_q(n;M_{\rm p}^q,B_{\rm p}^q).
\label{S10}
\end{equation}
In Eq.~(\ref{S10}), $ B_b^q $ gives the mean photon-(pair) number
per mode in component $b$ of TWB$_q$ composed of $ M_b^q $
independent identical modes, where $b$= p, s, and i in turn for
the photon-pair, noise-signal and noise-idler components and $ q =
{\rm p,a} $.

An iCCD camera as a photon-number-resolving detector is applied
for conditional photon-number resolving measurement(s) required
for the generation of the analyzed states. The camera is described
by a detection matrix $ T $ [specific for a given detection area
(DA) on its photocathode] that gives the probability of detecting
$ c $ photocounts when illuminated by $ n $ photons
\cite{PerinaJr2012}. Considering an iCCD camera with detection
efficiency $ \eta $, $ N $ equally-illuminated pixels, and $ d $
mean dark counts per pixel, we arrive at the formula
\begin{equation}  
 T(c,n;\eta,N,d) = \left(\begin{array}{c} N \\c \end{array}\right)
  (1-d)^N (1-\eta)^n (-1)^{c} 
  \sum_{l=0}^{c} \left(\begin{array}{c} c\\l\end{array}\right)
  \frac{(-1)^{l} }{ (1-d)^l }
  \left(1+\frac{l}{N}\frac{\eta}{1-\eta}\right)^n.
\label{S11}
\end{equation}

Following the scheme in Fig.~1 of the main text, we obtain the
following groups of states from the original twin beams TWB$ _{\rm
p} $ and TWB$_{\rm a} $. Different DAs with their corresponding
detection matrices are introduced as explained in Fig.~1 in the
main text.

\begin{description}
 \item[TS] A multi-mode thermal state (TS) is obtained as a marginal state of TWB$_{\rm p}$:
  \begin{equation}  
   p_{\rm th}(n_{\rm s}) = \sum_{n_{\rm i}=0}^{\infty}p^{\rm TWB}_{\rm p}(n_{\rm s},n_{\rm i}).
  \label{S12}
  \end{equation}

 \item[SPS] A sub-Poissonian state (SPS) with photon-number distribution
  $p_{\rm sP}$ [$p_{\rm a}$] arises in conditional photon-number resolving
  detection on the idler beam of TWB$_{\rm p}$ [TWB$_{\rm a}$]:
  \begin{eqnarray}  
    p_{\rm sP}(n_{\rm s};c_{\rm i}) &=& \sum_{n_{\rm i}=0}^{\infty}
    T_{\rm i}(c_{\rm i},n_{\rm i};\eta_{\rm i},N_{\rm i},d_{\rm i})  p^{\rm TWB}_{\rm p}(n_{\rm s},n_{\rm i}),\nonumber \\
    p_{\rm a}(n'_{\rm s};\bar{c}_{\rm a}) &=& \sum_{\bar{n}_{\rm a}=0}^{\infty}
   \bar{T}_{\rm a}(\bar{c}_{\rm a},\bar{n}_{\rm a};\bar{\eta}_{\rm a},\bar{N}_{\rm a},\bar{d}_{\rm a})
    p^{\rm TWB}_{\rm a}(n'_{\rm s},\bar{n}_{\rm a}).
  \label{S13}
  \end{eqnarray}
  The detection matrices $T_{\rm i}$ and $\bar{T}_{\rm a}$ belong
  to the DAs $ {\rm D}_{\rm i} $ and $ \bar{\rm D}_{\rm a} $,
  respectively. We note that the SPS with photon-number distribution $p_{\rm
  a}$ is used for PA.

  \item[PSTS] A state obtained by subtracting $ \bar{c}_{\rm s} $ photocounts ($ \bar{c}_{\rm s}$-PSTS) from
   the TS given in Eq.~(\ref{S12}) is obtained in the output of a beam splitter with transmissivity
   $t$ once $ \bar{c}_{\rm s} $ photocounts are detected in the
   other beam-splitter output monitored in $ \bar{\rm D}_{\rm s} $ with the detection matrix $\bar{T}_{\rm s}$:
   \begin{eqnarray}   
     p^{\rm s}_{\rm th}(n;\bar{c}_{\rm s}) &=& \sum_{\bar{n}_{\rm s}=0}^{\infty}
    \bar{T}_{\rm s}(\bar{c}_{\rm s},\bar{n}_{\rm s};\bar{\eta}_{\rm s},\bar{N}_{\rm s},\bar{d}_{\rm s})
     {\rm Bi}(n,n+ \bar{n}_{\rm s};t) p_{\rm th}(n+\bar{n}_{\rm s});
    \label{S14}
   \end{eqnarray}
   binomial distribution Bi is defined below Eq.~(\ref{S6}).

  \item[PATS] A state obtained by combining a sub-Poissonian field reached by post-selecting $\bar{c}_{\rm a}$
   photocounts in $ \bar{\rm D}_{\rm a} $
   ($\bar{c}_{\rm a}$-PATS) with a TS is described by the convolution of photon-number distributions given in Eqs.~(\ref{S12}) and (\ref{S13}):
   \begin{eqnarray}  
    p^{\rm a}_{\rm th}(n;\bar{c}_{\rm a}) &=& \sum_{n'_{\rm s}=0}^{n} p_{\rm th}(n-n'_{\rm s}) p^{\rm a}(n'_{\rm s};\bar{c}_{\rm a}).
    \label{S15}
   \end{eqnarray}

  \item[PSSPS] A state generated by subtracting $\bar{c}_{\rm s}$ photocounts, at a beam splitter
   with transmissivity $ t $ using DA $ \bar{\rm D}_{\rm s} $, from an SPS with
   photon-number distribution given given in Eq.~(\ref{S13})
   ($\bar{c}_{\rm s}$-PSSPS) is described by the formula:
   \begin{eqnarray}   
    p^{\rm s}_{\rm sP}(n;\bar{c}_{\rm s},c_{\rm i}) &=& \sum_{\bar{n}_{\rm s}=0}^{\infty}
    \bar{T}_{\rm s}(\bar{c}_{\rm s},\bar{n}_{\rm s};\bar{\eta}_{\rm s},\bar{N}_{\rm s},\bar{d}_{\rm s}) {\rm Bi}(n,n+ \bar{n}_{\rm s};t)
    p_{\rm sP}(n+\bar{n}_{\rm s};c_{\rm i}).
   \label{S16}
   \end{eqnarray}

  \item[PASPS] A state obtained by combining a sub-Poissonian field reached by post-selecting $\bar{c}_{\rm a}$
   photocounts in DA $ \bar{\rm D}_{\rm a} $
   with a SPS with photon-number distribution written in Eq.~(\ref{S13}) ($\bar{c}_{\rm
   a}$-PASPS) is characterized by the following photon-number
   distribution:
   \begin{eqnarray}  
    p^{\rm a}_{\rm sP}(n;\bar{c}_{\rm a},c_{\rm i}) &=& \sum_{n'_{\rm s}=0}^{n} p_{\rm sP}(n-n'_{\rm s};c_{\rm i})
    p^{\rm a}(n'_{\rm s};\bar{c}_{\rm a}).\nonumber \\
   \label{S17}
   \end{eqnarray}

  \item[PSTWB] Subtracting $\bar{c}_{\rm s}$ photocounts in $ \bar{\rm D}_{\rm s} $ belonging
   to the signal beam of twin beam TWB$ _{\rm p} $, we arrive
   at $\bar{c}_{\rm s}$-PSTWB with the following joint photon-number distribution:
   \begin{eqnarray}   
    p^{\rm s}_{\rm TWB}(n_{\rm s},n_{\rm i};\bar{c}_{\rm s}) &=& \sum_{n'=0}^{\infty}
    \bar{T}_{\rm s}(\bar{c}_{\rm s},\bar{n}_{\rm s};\bar{\eta}_{\rm s},\bar{N}_{\rm s},\bar{d}_{\rm s}) {\rm Bi}(n_{\rm s},n_{\rm s}+ \bar{n}_{\rm s};t)
    p_{\rm p}^{\rm TWB}(n_{\rm s}+\bar{n}_{\rm s},n_{\rm i}).
    \label{S18}
   \end{eqnarray}

  \item[PATWB] A state emerging by combining a sub-Poissonian field reached by post-selecting $\bar{c}_{\rm a}$
   photocounts in $ \bar{\rm D}_{\rm a} $
   with the signal beam of twin beam TWB$ _{\rm p} $ ($\bar{c}_{\rm
   a}$-PATWB) has the following joint photon-number distribution:
   \begin{eqnarray}  
    p^{\rm a}_{\rm TWB}(n_{\rm s},n_{\rm i};\bar{c}_{\rm a}) &=& \sum_{n'_{\rm s}=0}^{n_{\rm s}}
    p_{\rm p}^{\rm TWB}(n_{\rm s}-n'_{\rm s},n_{\rm i})
    p^{\rm a}(n'_{\rm s};\bar{c}_{\rm a}).\nonumber \\
    \label{S19}
   \end{eqnarray}
\end{description}

In the experiment, the values of parameters of both TWBs, as
described in Eq.~(\ref{S1}) and summarized in Tab.~\ref{tabS1},
were determined using the calibration method developed in
\cite{PerinaJr2012a,PerinaJr2013a} and applied to the
corresponding experimental joint photocount histograms. 
{We note that, in Tab.~\ref{tabS1}, there occur the Mandel-Rice distributions 
with the numbers $ M $ of modes considerably lower than 1. Such distributions
are highly peaked around the value $ n = 0$, which is a consequence
of the specific form of the noise occurring in the detection
process of an iCCD camera \cite{PerinaJr2017a}.} The values
of parameters characterizing different DAs on the photocathode of
the iCCD camera were obtained by this method (detection
efficiencies) and independent measurements: $\eta_{\rm
t}=\eta'_{\rm s}=\bar{\eta}_{\rm s} = 0.234\pm 0.005 $, $\eta_{\rm
i}=\bar{\eta}_{\rm a} = 0.227\pm 0.005 $, $ N_{\rm t} = N'_{\rm s}
= \bar{N}_{\rm s} =\bar{N}_{\rm a} =1512 $, $ N_{\rm i} = 3024$, $
d_{\rm t}=d'_{\rm s}=\bar{d}_{\rm s} =d_{\rm i}=\bar{d}_{\rm a}=
0.22/4536 $.
\begin{table}[t]    
 \centerline{
  \begin{tabular}{c c c c c c c }
   \hline
     { } \hspace{1mm} &  $ M_{\rm p} $ & $ B_{\rm p} $ & $ M_{\rm s} $ & $ B_{\rm s} $ & $ M_{\rm i} $ & $ B_{\rm i} $ \\
   \hline
     TWB$ {}_{\rm p} $ & 60.6 & 0.084 & 0.001 & 14.4 & 0.04 & 0.75 \\
     TWB$ {}_{\rm a} $ & 1740 & 0.001 & 0.23 & 1.28 & 0.34 & 0.94 \\
   \hline
  \end{tabular}    }
  \caption{{\bf Parameters of experimental twin beams
   TWB$_{\rm p} $ and TWB$_{\rm a} $}: Numbers $ M_b^q $ of modes with mean photon(-pair) numbers $ B_b^q $
   describing photon-pair component ($ b={\rm p} $), noise-signal component ($ b={\rm s} $),
   and noise-idler component ($ b={\rm i} $) of twin beam TWB$ {}_q $, $ q={\rm p,a} $. Relative errors are better than 10\%.}
\label{tabS1}
\end{table}

Both PS and PA are probabilistic processes because of
post-selection. We characterize them by the probabilities $ p^{\rm
p} $ of generation of the studied fields. These probabilities are
summarized in Table~\ref{tabS2} including both the experimental
values and the model predictions. In Tab.~\ref{tabS2}, as the used
TS emerges as the marginal state of twin beam TWB$_{\rm p}$ the
probabilities of success of PS and PA are the same for both kinds
of states. Also, the specific experimental conditions, that allow
to directly compare PS and PA, lead to close values of the
probabilities $ p^{\rm p} $ of generating the states $ \bar{c}
$-PSTWBs and $ \bar{c} $-PATWBs.

\begin{table}[htbp]  
\centering {\setlength{\tabcolsep}{4pt}
\begin{tabular}{l c c c c}
\hline
State    & $ \bar{c} =0$ & $ \bar{c} =1$ & $ \bar{c} =2$ & $ \bar{c} =3$  \\
\hline
PSTS/PSTWB & 50.10 (51.95) & 34.47 (33.76) & 11.98  (11.23) & 2.84  (2.55)  \\
PATS/PATWB & 51.49 (51.58) & 33.99 (33.84) & 11.35 (11.37) & 2.59 (2.63) \\
PSSPS & 46.69 (48.14) & 35.91 (35.59) & 13.42 (12.71) & 3.28 (2.96) \\
PASPS & 51.43 (51.58) & 34.08 (33.84) & 11.32 (11.37) & 2.60 (2.64) \\
PATSsc & 89.51 (89.49) &  9.79 (9.94)  &  0.65 (0.55) & - \\
PASPSsc  & 89.56 (89.50) &  9.78 (9.93)  & 0.66 (0.55) & - \\
\hline
\end{tabular}}
\caption{{\bf Probability $ p^{\rm p} $ (in \%) of generation of $
 \bar{c} $-PSTSs, $ \bar{c} $-PSTWBs, $\bar{c} $-PATSs, $\bar{c} $-PATWBs, $\bar{c} $-PSSPSs, $\bar{c}
 $-PASPSs, $\bar{c} $-PATSscs, and $\bar{c} $-PASPSscs}. The latter
 two correspond to $\bar{c} $-PATSs and $\bar{c} $-PASPSs,
 respectively, when spatial photon-pair correlations are taken into
 account. Theoretical predictions are given in parentheses and follow
 their experimental counterparts (errors $ \pm 1\%$).}
\label{tabS2}
\end{table}

\section{Quasi-distributions of intensities and their
reconstruction}

In a multi-mode optical field, a quasi-distribution $ P(W) $ of
field (integrated) intensity $ W $ related to the normal ordering
of field operators provides a complete description of its
statistical properties including all its nonclassical features
\cite{Mandel1995}. Knowing the quasi-distribution $ P(W) $ of
intensities, the Mandel detection formula \cite{Perina1991} gives
the corresponding photon-number distribution $ p $, that can be
compared with its reconstructed experimental counterpart:
\begin{equation}  
 p(n) = \frac{1}{n!}  \int_{0}^{\infty} dW \,
  W^{n} \exp(- W ) P(W).
\label{S20}
\end{equation}
The quasi-distribution $ P(W) $ of intensities in the normal
ordering of the field operators can be transformed into its
general $ s $-ordered form $ P(W;s) $ that gradually loses its
potential singularities and negativities as the ordering parameter
$ s $ drops down from $ s = 1$, which belongs to the normal
ordering \cite{Perina1991}.

Once the singularities of the quasi-distribution $ P(W;s) $ are
sufficiently suppressed by a suitable choice of the value of the
ordering parameter $ s $, we can reveal this quasi-distribution by
inverting the generalized Mandel detection formula
\cite{Perina1991}:
\begin{eqnarray} 
 P(W;s)&=& \frac{2}{(1-s)} \exp\left(-\frac{2 }{1-s} \right)
 \sum_{n=0}^{\infty}  \frac{p(n)}{n!}
  \left(\frac{s+1}{s-1}\right)^{n} 
  L_{n}\left( \frac{4W}{1-s^2}\right). \label{S21}
\end{eqnarray}
In Eq.~(\ref{S21}), symbols $ L_k $ denote the Laguerre
polynomials \cite{Morse1953}.

\begin{figure}[ht]   
\begin{centering}
  \centerline{ \includegraphics[width=0.25\hsize]{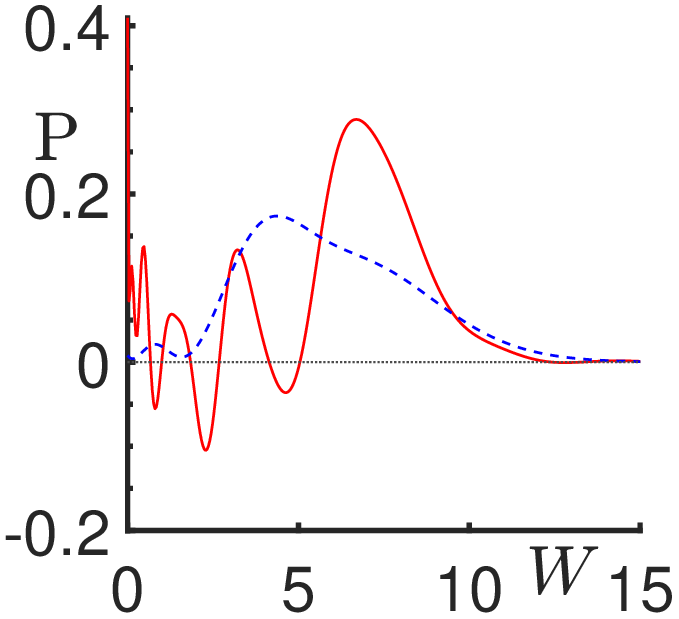}
   \includegraphics[width=0.37\hsize]{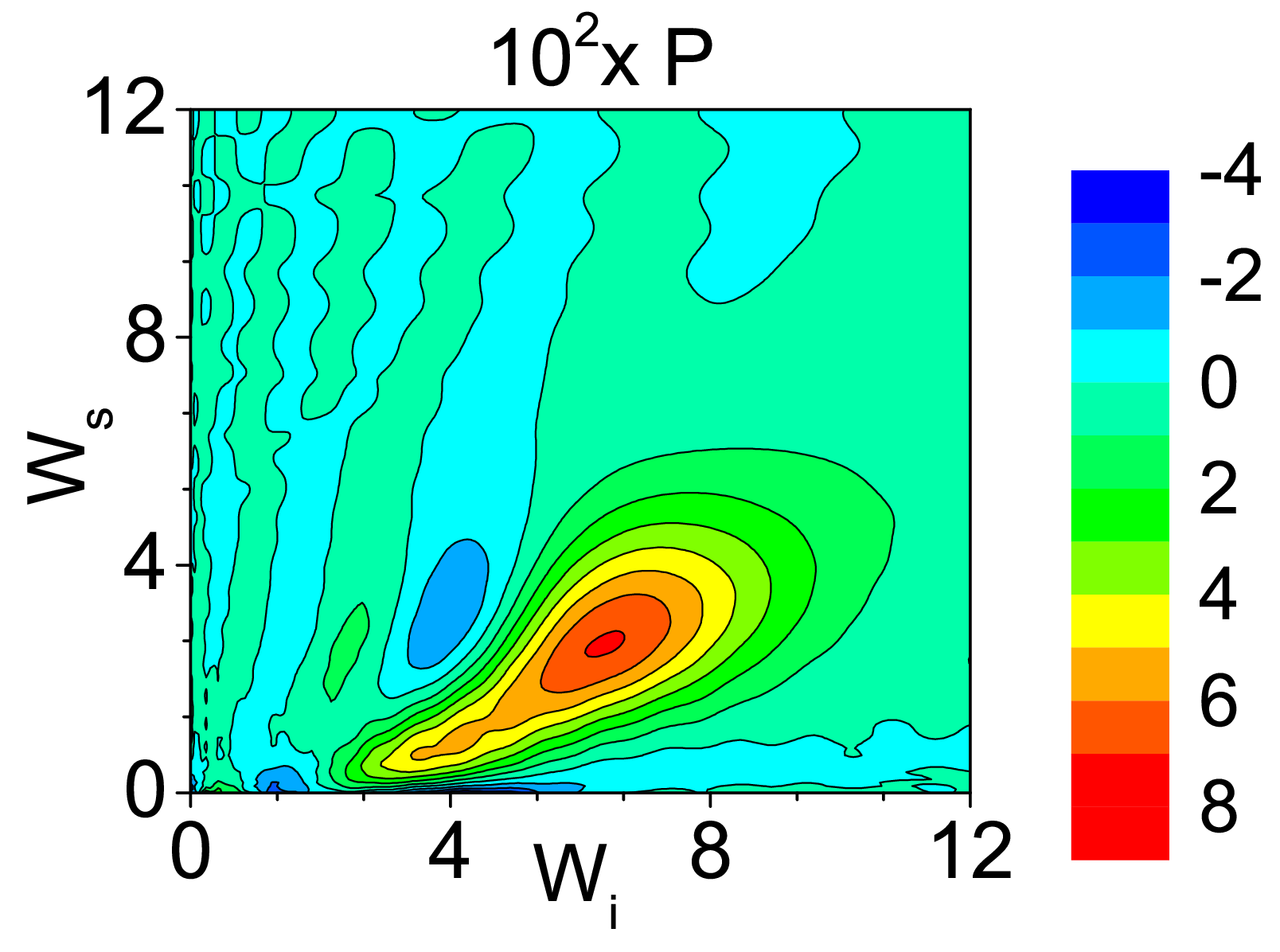}
   \includegraphics[width=0.37\hsize]{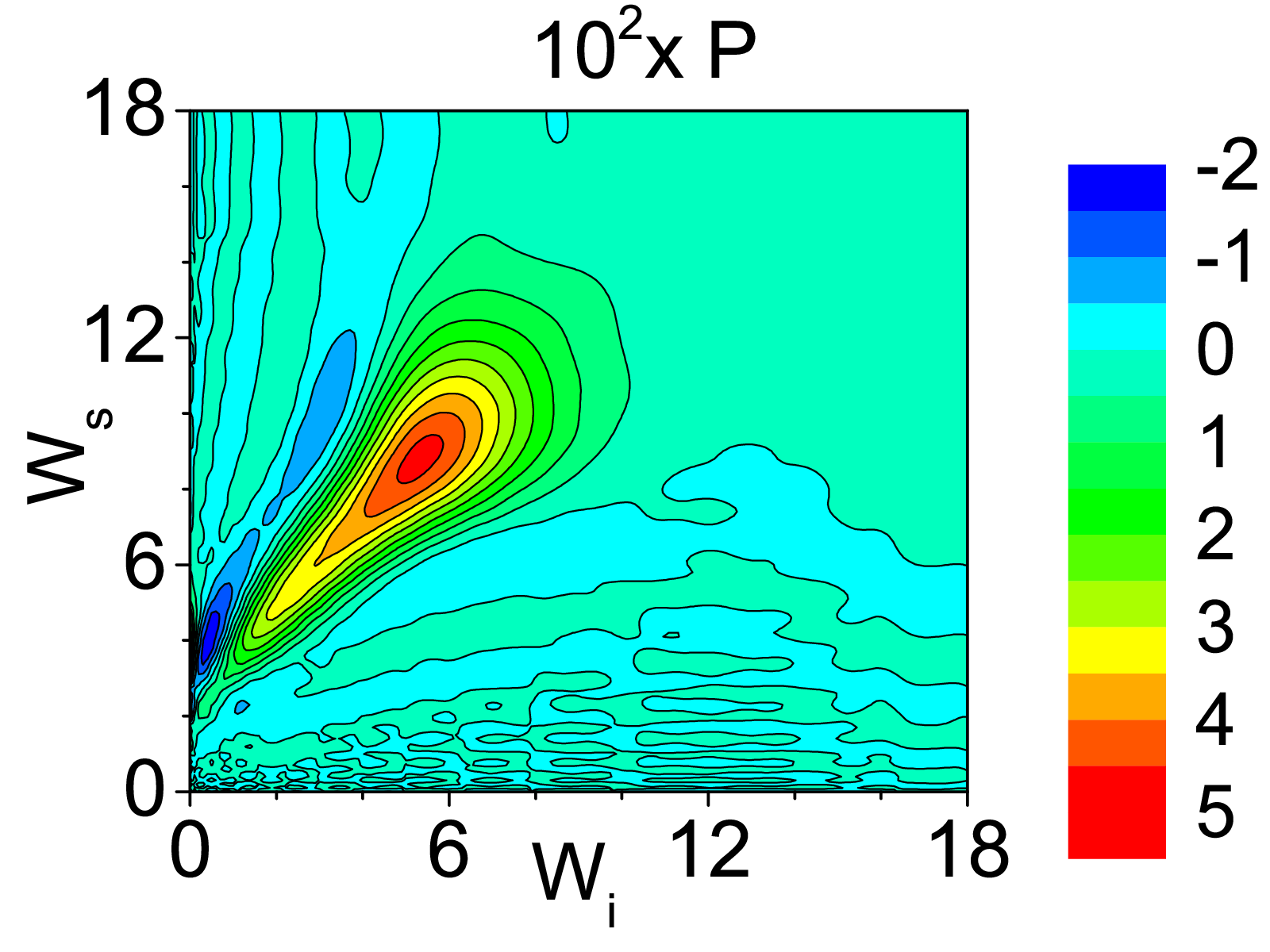}
    }
\centerline{ \small (a) \hspace{.3\hsize} (b) \hspace{.3\hsize} (c)}
 \par\end{centering}

 \caption{Quasi-distribution $P$ of intensities for (a) 2-PATSsc (blue
  dashed curve) and 2-PASPSsc (red curve) for $s=0.1$, (b) 2-PSTWB for
  $s=0.04 $, and (c) 2-PATWB for $s=0.04$.}
\label{figS2}
\end{figure}

The formulas in Eqs.~(\ref{S20}) and (\ref{S21}) can easily be
generalized to the fields composed of two beams
\cite{Haderka2005a,PerinaJr2017a,PerinaJr2022}. 
The one- and two-beam formulas then allow to compare the properties 
of PS and PA states from the point of view of their intensity
quasi-distributions. In these quasi-distributions, the nonclassicality 
of a state is manifested via their negative values
\cite{Glauber1963,Sudarshan1963}. The nonclassicality of the original 
SPS contrary to the classicality of TS results in greater
nonclassicalities of PASPSscs compared to PATSscs. This is
documented in Fig.~\ref{figS2}(a), where the quasi-distribution $P$
of intensity of 2-PASPSsc attains negative values while that of
2-PATSsc remains non-negative considering the same value of
ordering parameter $ s $. On the other hand, the above discussed
similarity of PS and PA in TWBs (exchanging the signal and idler
beams) is apparent also from the shapes of the corresponding
quasi-distributions $ P $, that are drawn in Figs.~\ref{figS2}(b,c)
for 2-PSTWB and 2-PATWB.

\bibliography{thapliyalSM}